\documentclass[twocolumn]{svjour3}          

\smartqed  
\usepackage{graphicx}
\usepackage{xspace} 
\usepackage[dvipsnames]{xcolor} 
\usepackage{siunitx}
\usepackage{wrapfig}
\usepackage{lineno}
\usepackage{amsmath}
\usepackage[titletoc,title]{appendix}
\RequirePackage{fix-cm}
\RequirePackage[colorlinks,citecolor=blue,urlcolor=blue,linkcolor=blue]{hyperref}

\definecolor{orcidlogocol}{HTML}{A6CE39}

\newcommand{\Hbb}{\ensuremath{H\to b\bar{b}}\xspace}
\newcommand{\Hcc}{\ensuremath{H\to c\bar{c}}\xspace}
\newcommand{\bjet}{\ensuremath{b}-\text{jet}\xspace}

\newcommand{\bjets}{\ensuremath{b}-\text{jets}\xspace}
\newcommand{\cjets}{\ensuremath{c}-\text{jets}\xspace}
\newcommand{\ljets}{\text{light-jets}\xspace}
\newcommand{\pt}{\ensuremath{p_\text{T}}\xspace}
\newcommand{\deltar}{\ensuremath{\Delta R(i,j)}\xspace}

\newcommand{\btagging}{\ensuremath{b}-\text{tagging}\xspace}
\newcommand{\btagged}{\ensuremath{b}-\text{tagged}\xspace}
\newcommand{\WZjets}{\ensuremath{W/Z}+\text{jets}\xspace}
\newcommand{\tteff}{\ensuremath{\tilde{\epsilon}}\xspace}

\newcommand{\nnoutputindex}{\ensuremath{\epsilon}_{{NN}}(\mathbf{\Theta}_e)_i\xspace}
\newcommand{\eventeff}{\ensuremath{\epsilon_{\text{event}}}\xspace}
\newcommand{\jeteff}{\ensuremath{\epsilon_{\text{jet}}}\xspace}

\usepackage{etoolbox}
\newcommand*{\affaddr}[1]{#1} 
\newcommand*{\affmark}[1][*]{\textsuperscript{#1}}
\begin{document}

\title{Efficiency Parameterization with Neural Networks}


\author{%
 Francesco Armando Di Bello\affmark[1,2] \and Jonathan Shlomi\affmark[3] \and Chiara Badiali\affmark[1,2] \and Guglielmo Frattari\affmark[1,2] \and Eilam Gross\affmark[3] \and Valerio Ippolito\affmark[2] \and  Marumi Kado\affmark[1,2, 4] 
}
\authorrunning{F.A. Di Bello, J. Shlomi \and  et al.}

\institute{
              Francesco Armando Di Bello \\
              \email{Francesco.Armando.DiBello@roma1.infn.it} \\  
              \and
              Jonathan Shlomi \\
              \email{Jonathan.shlomi@weizmann.ac.il}\\
             \at
\affaddr{\affmark[1]Dipartimento di Fisica, Sapienza Universit\` a di Roma, Roma, Italy}\\
\affaddr{\affmark[2]INFN Sezione di Roma, Roma, Italy}\\
\affaddr{\affmark[3]Department of Particle Physics, The Weizmann Institute of Science, Rehovot, Israel}\\
\affaddr{\affmark[4]Universit\' e Paris-Saclay, CNRS/IN2P3, IJCLab, 91405, Orsay, France}\\
}

\maketitle

\begin{abstract}
Multidimensional efficiency maps are commonly used in high energy physics experiments to mitigate the limitations in the generation of large samples of simulated events. Binned efficiency maps are however strongly limited by statistics. We propose a neural network approach to learn ratios of local densities to estimate in an optimal fashion efficiencies as a function of a set of parameters. Graph neural network techniques are used to account for the high dimensional correlations between different physics objects in the event. We show in a specific toy model how this method is applicable to produce accurate multidimensional efficiency maps for heavy flavor tagging classifiers in HEP experiments, including for processes on which it was not trained.
\end{abstract}

\section{Introduction}
\label{sec:intro}

An overarching issue of Large Hadron Collider (LHC) experiments is the necessity of massive numbers of simulated collision events to estimate the rates of expected processes in very restricted regions of phase space. To mitigate this difficulty, a commonly used approach is the \emph{event weighting technique} which replace selection cuts with event weights. These weights are typically defined from binned efficiency maps. The difficulty in these methods is the range of applicability of efficiency maps that are limited in the number of dimensions (typically two), and subsequently, fail to capture more subtle effects that appear in specific regions of phase space. To account for these dependencies,  a multidimensional mapping is required. This implies large statistical fluctuations in the map itself that defies the original purpose of the method.

A common example of the usage of event weighting techniques is typically given by analyses relying on the identification  of jets originating from $b$-quarks ($b$-tagging) \cite{btaggingATLAS,VHcc,HbbATLAS}. Applying a weight corresponding to the expected identification efficiency of a jet, i.e.  the probability of being identified as a $b$-jets,  instead of a direct selection cut can provide large gains in statistics (especially in cases of percent level efficiencies to be applied on several jets in an event). However, obtaining universally applicable maps require to account for a large number of parameters. Some of which  are typically not known.

The goal of the proposed method is to provide higher dimensional parametrizations of efficiencies that can capture non-trivial dependencies while making optimal use of the available statistics and therefore be applicable in any analysis context considered. When achieving this goal the parameterization will be referred to as {\it universal}. The proposed approach is based on Graph Neural Networks (GNN). The case study used is the $b$-tagging performance in the analysis of Higgs boson decays to $b$-quarks (\Hbb).

The strength of the proposed method relies on its ability to model high-dimensional correlations between jets. These jet-by-jet dependencies are not given explicitly as inputs variables to the neural network, but rather they are inferred  from single-jet properties  during the training of the network. In case multiple jets in the event are $b$-tagged, the jet-efficiencies provided by the NN can be combined to derive an unbiased estimator of the event tagging efficiency. A toy model is built to probe the capability of the ML approach to provide a robust parametrization of the \btagging efficiency.

The paper is organized as follows.  Section \ref{sec:eventW} introduces the event weighting technique and describes the main challenges and goals of the method. Section \ref{sec:samples} describes the MC simulation technique used to generate the toy data-set. Section \ref{sec:truthtagging} describes a map-based technique that is commonly used to estimate the event weight based on a parameterization of the \btagging classifier performance. Section \ref{sec:GNN} describes the GNN model, whose results are compared to the ones of the map-based technique in Section \ref{sec:res}. In Section \ref{sec:discussions}  some considerations about the usage of the proposed methodology in real experiments are presented. Conclusions are drawn in Section \ref{sec:conclusions}.

\section{Event weighting technique}
\label{sec:eventW}

In high energy physics experiments (HEP), estimating a background rate or a signal efficiency from a selection cut is most accurately achieved by a full simulation of the event. However, the  precision of such an estimate can be heavily affected by the limitation in the number of events that can be simulated in a given region of phase space. If instead of selecting events based on a classification cut, a weight corresponding to the classifier efficiency is applied, significant improvements in sensitivity can be gained. This procedure is also known as \emph{Tag-Rate-Function (TRF) method} or \emph{Truth Tagging (TT)}\cite{truthtagging,truthtagging_thesis}.  


Selections can be interpreted as a classification depending on a vector of input variables $\mathbf{x}$. The classifier can be represented by a function $f(\mathbf{x})$ and the classification by a simple selection cut on the classifier above a given threshold $T_f$. The classifier can represent simple cuts or a multivariate method. Typically the variables $\mathbf{x}$ depend on several underlying variables which will be denoted  by $\boldsymbol{\theta}$.

\begin{figure*}

\centering

    \includegraphics[]{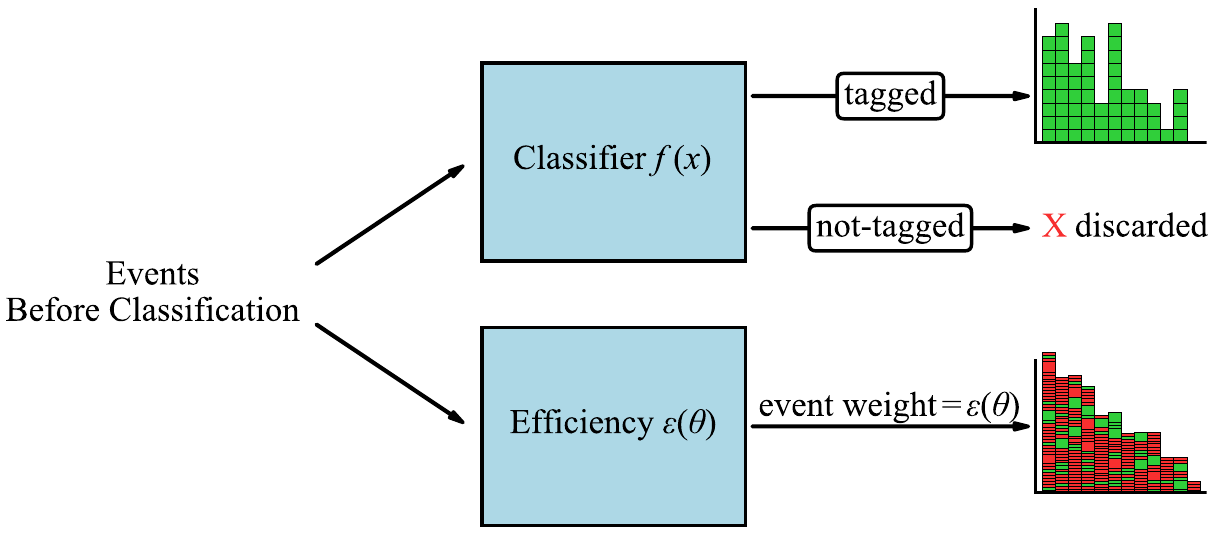}
        \caption{Usage of event weighting to reduce MC statistical uncertainties of some observable distribution. The plot on the top shows a classifier $f(\mathbf{x})$ used to select events. The events which pass the classification requirement are represented in green while the rejected events are shown in red. The bottom panel shows the event weighting  where the classifier efficiency $\epsilon(\boldsymbol{\theta})$ is used to weight the events rather than rejecting them. $\mathbf{x}$ are the variables used by the classifier. For \btagging, $\mathbf{x}$ includes variables such as the secondary vertex information while $\boldsymbol{\theta}$ is the set of relevant variables used for the parametrization of the efficiency, such as the jet $\pt$ and $\eta$.}
    \label{fig:truthtag_explain}

\end{figure*}

In the case of Heavy Flavor tagging, $\boldsymbol{\theta}$ is typically defined as the jet transverse momentum $\pt$ and pseudo-rapidity $\eta$~\cite{btaggingATLAS}, while $\mathbf{x}$ includes the reconstruction of secondary vertices and a combination of track impact parameter information estimated from the properties of a set of reconstructed charged particle tracks. This information is then combined to produce a multivariate jet-based classifier $f(\mathbf{x})$. Figure~\ref{fig:truthtag_explain} schematically shows the usage of the efficiency for event weighting to reduce MC statistical uncertainties.

A parametrized classifier efficiency can be written as:

\begin{equation}
\epsilon_{\textrm{jet}}\left(\boldsymbol{\theta}\right) = \frac{N(f(\mathbf{x}) > \textrm{T}_f| {\boldsymbol{\theta}})}{N(\boldsymbol{\theta}) } 
\end{equation}
where $T_f$ is the operating working point threshold of the classifier; the numerator, the selected number of jets of a given flavor at this working point; and the denominator represents the total number of jets of the same flavor. 

\begin{figure*}
\centering   
    \includegraphics[]{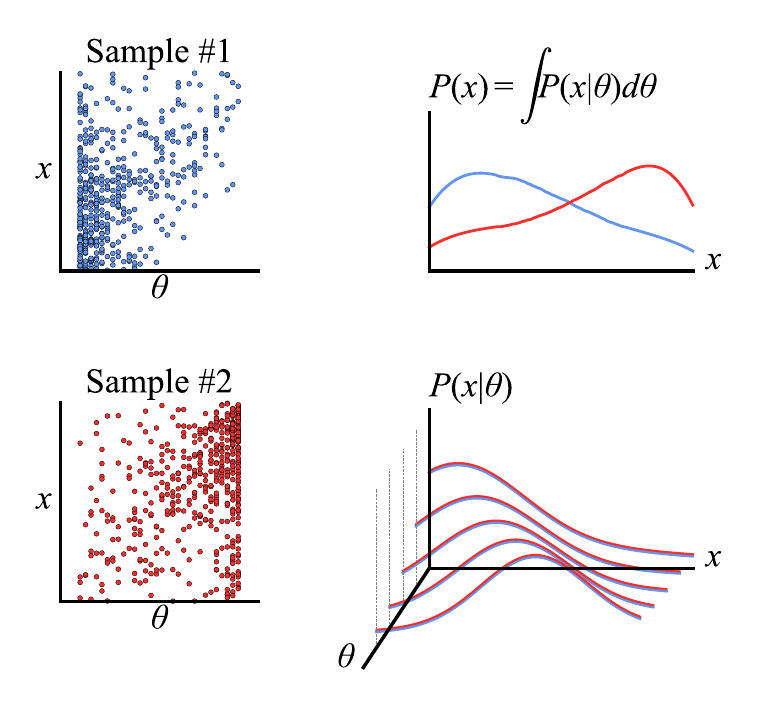}

        \caption{Illustration of a universal parametrization of the classifier efficiency. The joint distribution of  ($\mathbf{x},\boldsymbol{\theta}$) is generally different between two samples. The top right plot  shows  the overall probability distribution of the  input variables of the classifier, $P(\boldsymbol{x})$, for two different samples. Different $P(\mathbf{x})$ distributions lead to  different overall efficiencies between the two samples. The bottom right plot shows the conditional probability distributions, $P(\mathbf{x}|\boldsymbol{\theta})$, between the two samples. The set of relevant variables $\boldsymbol{\theta}$ is defined to provide a $P(\mathbf{x}|\boldsymbol{\theta})$ which is sample independent. Under this condition, the parametrized classifier efficiency $\epsilon(\boldsymbol{\theta})$ is expected to be universal.}
    \label{fig:parametrized_eff_explain}
\end{figure*}

To achieve a parametrization of the efficiency, applicable to a large number of analyses, a set of relevant variables $\boldsymbol{\theta}$ must be defined such that the conditional probability of the classifier inputs, $x$, at a given value of $\boldsymbol{\theta}$, $p(\mathbf{x}|\boldsymbol{\theta})$, will be identical between samples or different regions of phase space, as illustrated in Figure~\ref{fig:parametrized_eff_explain}. 

This motivates the \emph{efficiency maps} approach, where an attempt is made to parametrize $\epsilon_{\textrm{jet}}$ binned in $\boldsymbol{\theta}$. Efficiency maps are a commonly used tool in real experiments. However, taking into account the full dependencies of the classifier efficiency is often impractical using efficiency maps. The reason being that a small enough set of variables that fully capture these dependencies might not be available. 

In the case of \btagging it was found that while $p_{T}$ and $\eta$ are indeed the most dominant variables in determining $\jeteff$, there are other variables that affect the efficiency and could be considered had we known them,  e.g. the angular separation and flavor of the adjacent jets~\cite{HBBRUN1,VHcc}.

We propose a different approach to estimate $\jeteff$ based on a neural network built using a GNN. The neural network takes as input a set of jet-variables $\mathbf{\Theta}_{j_{e}}$ for each jet $j$ in the event $e$. The input variables are the  jet-($\pt$, $\eta$, $\phi$, $\text{flavor}$) and the neural network model infers, in addition to $\pt$ and $\eta$,  the jet-by-jet angular dependencies  of the \btagging efficiency which reflects the environment of the $b$-tagged jet.

\section{Simulated samples}
\label{sec:samples}
The samples employed in this study consist of toy $pp$ collision events  with multiple jets generated with generic kinematic and flavor properties. 
We assume a cylindrical coordinate system where particle beams collide on the $z$ axis, $xy$ is denoted as the transverse plane, $\phi$ is the azimuthal angle, ${\theta}$ the polar angle, and pseudo-rapidity $\eta$ is defined as $\eta=-\log\tan(\theta/2)$.


The generated events are sampled using an exponential function to fix the number of jets in the event and Gaussians or polynomial distributions to sample the jet kinematics variables and the angular distance between two jets $\deltar=\sqrt{(\eta_i-\eta_j)^2 + (\phi_i-\phi_j)^2}$. More details about the event generation can be found in Appendix \ref{app:samp}. 

\begin{figure*}
\centering
    \includegraphics[]{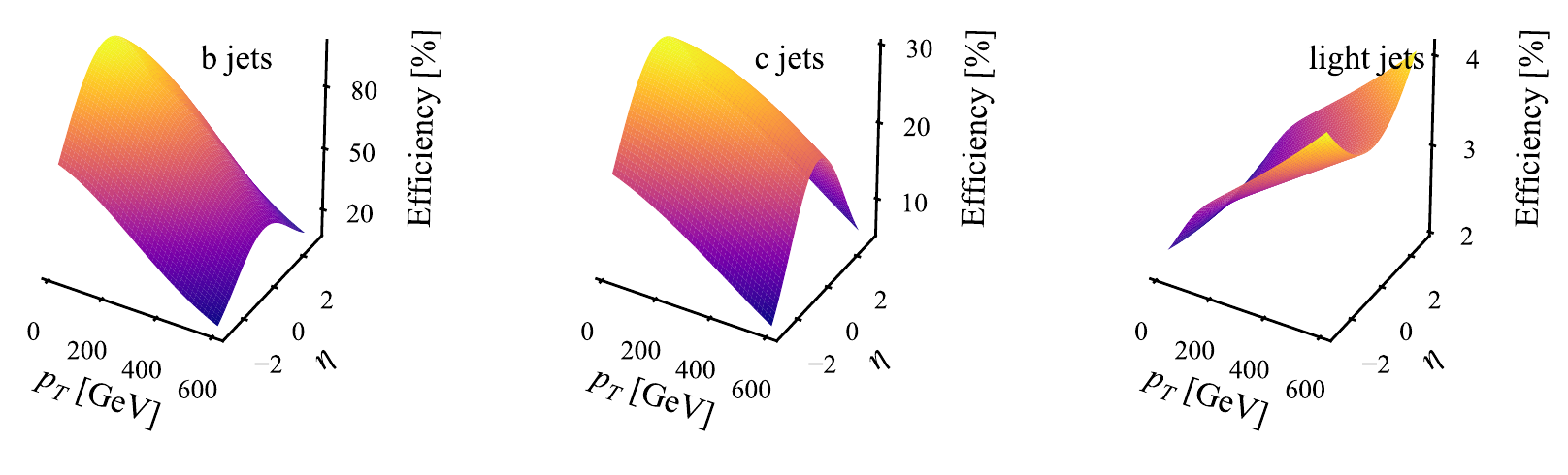}
    \includegraphics[]{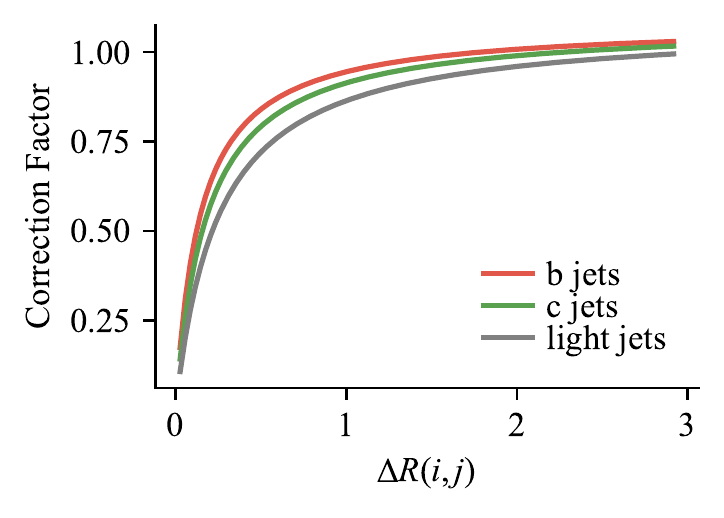}

\caption{The parameterized efficiencies used to emulate the performance of the flavor tagging algorithms.  (a) The efficiencies for each flavor as a function of $p_{T}$ and $\eta$, $\epsilon_{f_i}(\pt,\eta)$. (b) The multiplicative correction factor  $\hat{\epsilon}_{ij}\left(\Delta R(i,j), f_j \right)$ which accounts for the proximity ($\deltar$) and flavor of the close-by-jet $f_j$.}
\label{fig:maps}
\end{figure*}

Three separate samples of four-momenta representing $b$-, $c$- and light-jets are generated. The \btagging efficiency is modeled using ad-hoc parameterizations using a multivariate Gaussian distribution depending on  $\pt$ and $\eta$ which is modified by a multiplicative correction factor depending on the angular distance $\deltar$ of other jets in the event as well as their flavor. This efficiency is chosen to mimic the \btagging performance of ATLAS and CMS\cite{btaggingATLAS,btaggingCMS} and it is expressed as:
\begin{equation}
\centering
\label{eq:trueeff_jet}
 \hspace{.1\linewidth}  
{\jeteff}_{i} = \epsilon_{f_{i}}(\pt,\eta) \cdot \prod_{j} \hat{\epsilon}_{ij} \left(\Delta R(i,j), f_j \right),
\end{equation}
where $\epsilon_{f_{i}}(\pt,\eta)$ is the two-dimensional parameterisation of the efficiency to tag a jet of a given flavor $f_i$, and  $\hat{\epsilon}_{ij}\left(\Delta R(i,j), f_j \right)$  is the one-dimensional correction factor which accounts for the effect of any close-by jet $j$ of flavor $f_j$ in the event. The efficiencies $\epsilon_{f_i}(\pt,\eta)$ and  the correction factors $\hat{\epsilon}_{ij}\left(\Delta R(i,j), f_j \right)$ are shown in Figure~\ref{fig:maps}.

The true \btagging  efficiency of each individual jet in the event is computed using Eq. \ref{eq:trueeff_jet}. This efficiency value ${\jeteff}_i$ is used to emulate \btagging by assigning a boolean value to each jet $\texttt{istag}$ which is set to 1 based on a random score $s_i$ sampled from a uniform distribution. Namely, if $s_i < {\jeteff}_{i}$ the $i$-th jet in the event is considered to be $b$-tagged ($\texttt{istag}$=1).
In many physics analyses, multiple jets in the event are required to pass \btagging selections, hence the efficiencies of the single jet need to be combined to form a per-event efficiency. In this toy analysis the event selection is based on the two jets with highest \pt in the event (``leading jets'', labeled as $1$ and $2$), and it is defined depending on the number of $b$-tagged jets, $n_\textrm{tag}$:
\begin{equation}\label{eq:trueeff_event}
    \eventeff = 
        \begin{cases}
        \quad(1-\epsilon_1) (1-\epsilon_2)&\qquad\text{if $n_\textrm{tag}=0$},\\
        \quad\epsilon_1 (1-\epsilon_2)+(1-\epsilon_1) \epsilon_2&\qquad\text{if $n_\textrm{tag}=1$},\\
        \quad\epsilon_1 \epsilon_2&\qquad\text{if $n_\textrm{tag}=2$}.
        \end{cases}
\end{equation}

\section{Efficiency Map techniques}
\label{sec:truthtagging}
The estimation of $\eventeff$ in the case of \btagging in real experiments is commonly based on the binned two-dimensional efficiency maps in the jet \pt-$\eta$ plane \cite{truthtagging,truthtagging_thesis}, $\tteff$, derived from MC simulation separately for \bjets, \cjets and \ljets, which are used to approximate the per-jet \btagging efficiency of Eq. \ref{eq:trueeff_jet}  as:
\begin{equation}\label{eq:truthtagging}
\jeteff \approx \tteff_i = \tteff_{f_i}(\pt,\eta).
\end{equation}
The choice of the variables used to parameterize \tteff is motivated  by the expected dependency of the \btagging performance. For example, as the transverse momentum of a \bjet increases, the dilation of its lifetime in the laboratory frame results in  secondary decay vertices which are reconstructed further from the interaction point of the primary collision. The reconstruction efficiency of secondary vertices is not constant as a function of their distance to the primary vertex and this affects the response of the \btagging classifier. Similarly, the typical configuration of multi-purpose detectors produces a dependency of track reconstruction performance on detector geometry, which in turn propagates into a dependency of the \btagging performance on $\eta$.

From the per-jet efficiency maps \tteff the event weight \eventeff is computed factorizing the contribution from the various jets, similarly to what is shown in Eq. \ref{eq:trueeff_event}.

The main limitation of this map-based approach is the assumption that correlations between jets can be neglected and that the efficiency of \btagging a single jet only depends on its \pt and $\eta$. The dependency of efficiency on residual observables is marginalized out when deriving $\tteff$ from MC samples, introducing a bias that is particularly significant for final states with large jet multiplicities or events where close-by or overlapping jets are reconstructed from the decay of boosted resonances. 
A dedicated $\deltar$ reweighing was derived and used to correct for this effect in previous \Hbb and \Hcc analyses\cite{HBBRUN1,VHcc}. Given the uncertain nature of this correction and the limited statistics of the sample used to derive it, a large systematic uncertainty equal to half of the correction was assigned to the relevant MC templates. The overall uncertainty related to the statistics of the MC templates  constitutes a contribution up to around  20\% to the total background uncertainty \cite{HbbATLAS,HbbCMS}.  

Additional limitations come from the binning of the two-dimensional maps. To reduce discontinuities, smoothing techniques need to be employed. However, these techniques often require a non-trivial interplay between the bin sizes and the parameters of the smoothing model resulting unpractical compared to an unbinned neural network training. Finally, the NN technique provides a simultaneous estimate of the efficiency for each jet-flavor in contrast to the map-based approach which requires a dedicated parametrization for each of the flavor independently.

\section{Truth Tagging with Neural Networks}
\label{sec:GNN}
Taking into account the full dependency of the jet-tagging probability on all event observables would be unpractical with a map-based approach. ML techniques, on the other hand, provide the possibility to scale the problem to higher dimensionality and therefore to more challenging physics topologies.

In principle, a standard feedforward neural network could be used to perform the task. However, these models are not able to optimally cope with inputs of variable sizes and thus the overall correlations between jets in the event cannot be easily exploited during the training.  The technique we propose uses a GNN to capture efficiently these correlations.
A GNN also offers a more natural representation of the data by exploiting  pair-wise relationships between the jets. In our toy experiment, each jet is represented by a set of variables corresponding to $(\pt,\eta,\phi,\text{flavor})$. The neural network takes as input these variables for each jet in the event $e$, $\mathbf{\Theta}_{e}$ = $(({\pt}_{1},{\eta}_{1},{\phi}_{1},\textrm{flavor}_{1})$, ..., $({\pt}_{n_{jets}}$, ${\eta}_{n_{jets}}$, ${\phi}_{n_{jets}}$,$\textrm{flavor}_{n_{jets}})$) and learns to approximate the efficiency given in Eq. \ref{eq:trueeff_jet} for each of these jets. Note that the inputs to the neural network do not include $\Delta R$ between neighboring jets, which is the variable that determines the correction applied in Eq. \ref{eq:trueeff_jet} but rather this dependency is inferred directly during the training. 

\begin{figure*}

    \centering
    \includegraphics[]{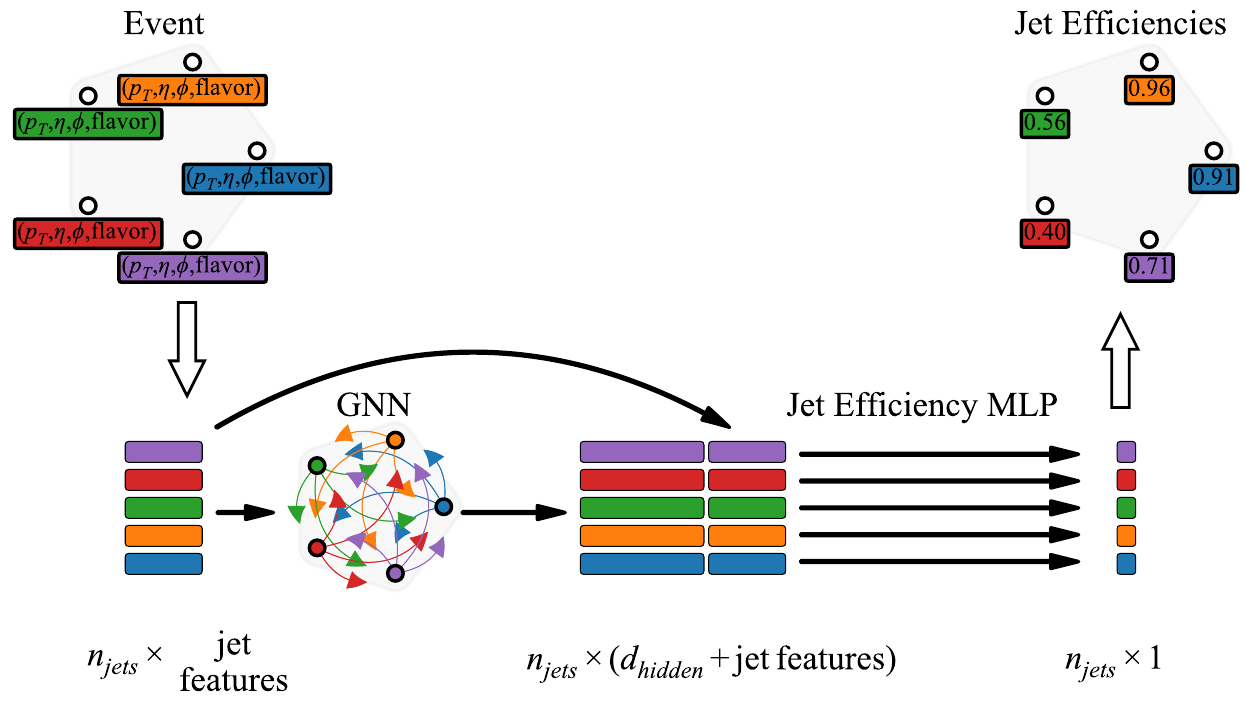}

    \caption{Schematic representation of the neural network structure.}
    \label{fig:nn_arch}
\end{figure*}

\paragraph{Model Architecture}
The model, referred to as NN in the following, consists of two components: a GNN\cite{GNN2} and a \emph{jet efficiency} network. The flow of information between the different parts is illustrated in Figure \ref{fig:nn_arch}.

The GNN component creates a \emph{hidden representation} for each jet that is based on the information of the other jets in the event. The GNN takes as input the $n_{jets}\times 4$ matrix of jet features, and outputs  $n_{jets}\times d_{hidden}$  matrix of jet hidden representations\footnote{$d_{hidden}$, a hyperparameter of the model, is the size of this representation.}.
The jet efficiency network then operates on each jet individually. It takes as an input the jet variables and the jet hidden representation and it returns as an output the \jeteff for every jet. More details about the model architecture can be found in Appendix~\ref{app:MA}.

\paragraph{Training Procedure}
The network is trained to predict the $n_{jets}\times 1$ vector of efficiencies. The loss function used for training is the weighted binary cross-entropy (BCE), which for a single event it can be written as:

\begin{figure*}
   \centering
     \includegraphics[width=.8 \textwidth]{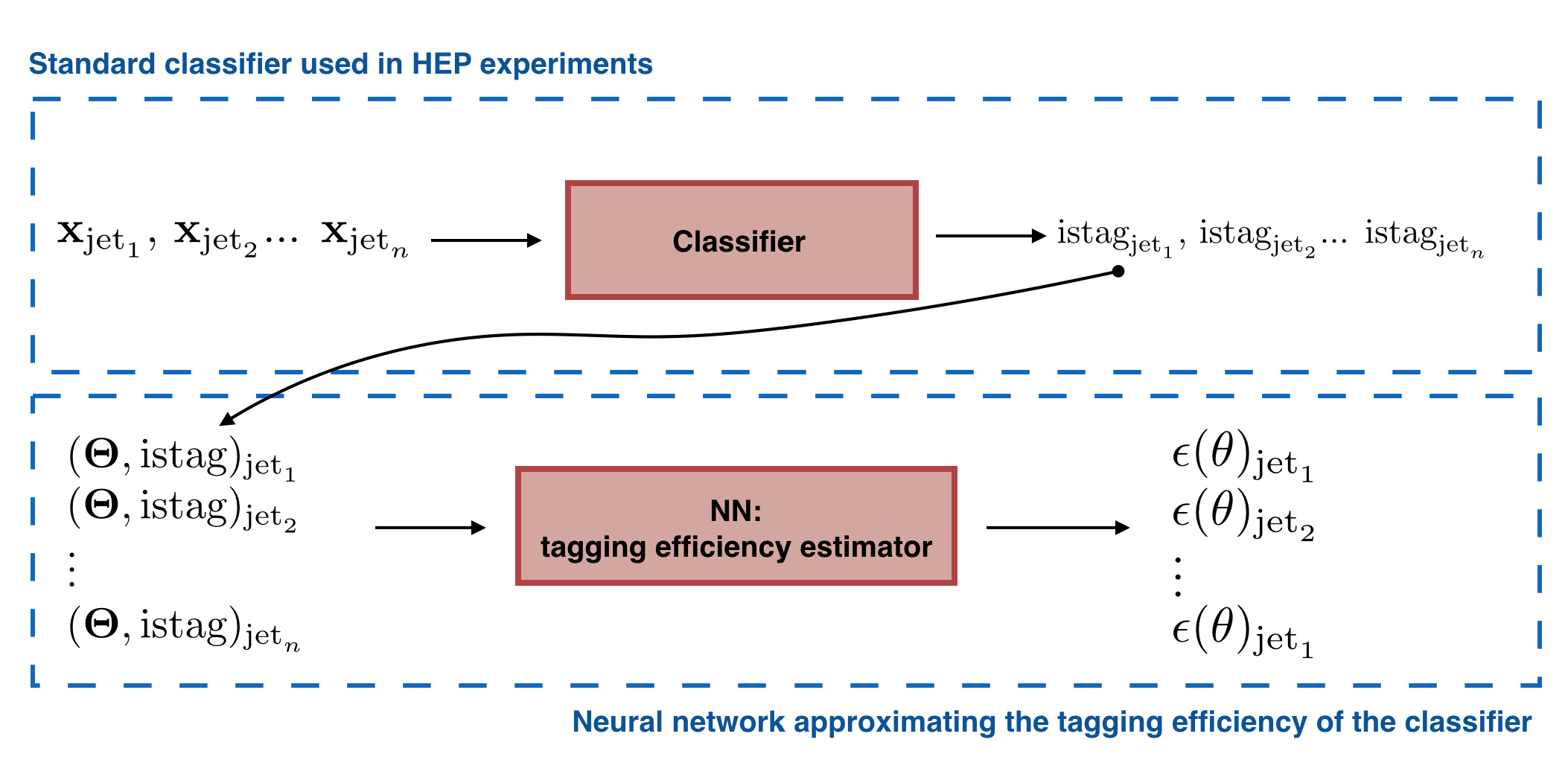}
        \caption{Schematic representation of the proposed training workflow. In the top box a standard classifier, such as the one used for \btagging, is represented. The classifier takes as input variables a set of jet observables, $\mathbf{x}_{\text{jet}_i}$, for each jet $i$. An output boolean, ``istag'', is associated to each jet, representing whether the jet is classified as a \bjet or not. In typical applications of \btagging in HEP experiments, $\mathbf{x}_{{\text{jet}}_{i}}$ represents, among other variables, the properties of reconstructed secondary vertices in the jet. The NN acts as an efficiency estimator, it takes as an input a different set of variables describing all jets in the event to which \btagging is applied, ${\mathbf{\Theta}_e}=({\mathbf{\Theta}_1}, {\mathbf{\Theta}_2}$...${\mathbf{\Theta}_n})$ where ${\mathbf{\Theta}_i}=({\pt}_{i},{\eta}_{i},{\phi}_{i},\text{flavor}_{i})$, and the output of the classifier, $\texttt{istag}_{i}$, is used to label the classes given to the NN. The NN treats all input jets in the event simultaneously, so those correlations are taken into account and $\epsilon(\boldsymbol{\theta}_{\textrm{jet}})$ can be estimated for each jet in the event.}
    \label{fig:whatarewedoing}
\end{figure*}

\begin{equation}
\begin{split}
\label{loss}
\text{BCE}_{e} = \frac{1}{N_{jets}}\sum^{N_{jets}}_{i} \left[- ({\texttt{istag}_{i}}) \log(\nnoutputindex)\right]\\
- \left[ \mu  (1-{\texttt{istag}}_{i}) \log(1-\nnoutputindex)\right],
\end{split}
\end{equation}

where the sum runs over the sets of jets, $N_{jets}$, in the event, $e$, which pass ($\texttt{istag}$=1) and do not pass ($\texttt{istag}$=0) \btagging and $\nnoutputindex$ is the $i$-th component of the output of the NN, a vector of variable size representing the predicted efficiency of tagging each jet in an event. The loss function being minimized is the sum of $\text{BCE}_e$ for all the events in the training sample. The factor $\mu$ controls the weight of the non-tagged events and can be used to balance the number of tagged and non-tagged jets to facilitate the training. 
 This approach could be useful for light-jets where the number of non-tagged jets is $\mathcal{O}(100)$ larger than the tagged ones. Even if this factor was found to be helpful in tests conducted with feedforward networks, for GNNs it was found to have a negligible impact on the final results. Therefore, $\mu$=1 is assumed in the following discussions.
 
Using a well-known result, the neural network trained using BCE as loss function converges to the following ratio~\cite{FER,losstrick}:
\begin{equation}
\begin{split}
\label{eq:pratio}
\nnoutputindex \approx \frac{p_{\text{tag}}(g_i(\mathbf{\Theta}_e))}{p_{\text{tag}}(g_i(\mathbf{\Theta}_e)) + p_{\text{non-tag}}(g_i(\mathbf{\Theta}_e))}\\
 \approx \epsilon(g_i(\mathbf{\Theta}_e)) \approx \epsilon({\boldsymbol{\theta}_i}) = \epsilon_i
\end{split}
\end{equation}
 $\nnoutputindex$ and  $\epsilon({\boldsymbol{\theta}}_{i})$ are the predicted and true efficiency jet of the $i$-th jet in the event $e$, respectively. In the toy model employed for this study, $\epsilon({\boldsymbol{\theta}}_{i})$ represents the true single-jet efficiency, $\epsilon_i$, computed in Eq.~\ref{eq:trueeff_jet}. $g_i(\mathbf{\Theta}_e)$ is the function, infeered during the training, which approximate the relevant variables of the $i$-th jet $\boldsymbol{\theta}_i$, $g_i(\mathbf{\Theta}_e) \approx \boldsymbol{\theta}_i$. For example, for the $i$-th jet in the event: $g_i  ( ({\pt}_{1},{\eta}_{1},{\phi}_{1},{\text{flavor}}_{1})$, ...,  $({\pt}_{n_{jets}}, {\eta}_{n_{jets}}$,${\phi}_{n_{jets}}$, ${\text{flavor}}_{n_{jets}} ))$  $\approx ({\pt}_{i},{\eta}_{i},\deltar,{\text{flavor}}_{j})$  where the index $j$ runs over every jet in the event, excluding the  $i$-th jet. Finally, $p_{\text{non-tag}}(g_i (\mathbf{\Theta}_e))$ and $p_{\text{tag}}(g_i (\mathbf{\Theta}_e))$ are the $g_i(\mathbf{\Theta}_e)$ distributions of the $i$-th jet to be non-tagged and tagged as a $b$-jet, respectively.

It is worth noticing that the NN computes directly the efficiency $\nnoutputindex$ without regressing $p_{\text{tag}}(g_i(\mathbf{\Theta}_e))$ and $p_{\text{non-tag}}(g_i (\mathbf{\Theta}_e))$ independently. In the map-based approach, on the other hand,  the distribution of tagged jets, $p_{\text{tag}}({\pt}_i,{\eta}_i)$, and the distribution of the total number of jets, which is the sum of tagged and non-tagged jets ($p_{\text{tag}}({\pt}_i,{\eta}_i)+p_{\text{non-tag}}({\pt}_i,{\eta}_i)$), are computed independently in bins of $\pt$ and $\eta$. The efficiency is then estimated in a second step by taking the ratio bin-by-bin between these two distributions.

The training workflow of the proposed approach is illustrated in Figure \ref{fig:whatarewedoing}.

The training is done with stochastic gradient descent, with a batch size of 5,000 events. The batch size was chosen to be as large as possible given the memory constraints of the system used for training. The batch size is particularly important for this task as a significant amount of tagged and non-tagged jets needs to be present to reduce statistical fluctuations during training. To further reduce the effect of these fluctuations, 20 neural networks with different weights initialization and random batching during training were used. The efficiency for each jet is computed by taking the mean of these 20 different predictions.

\section{Results}
\label{sec:res}
In this section, the result of approximating $\eventeff$ using the jet \btagging efficiencies calculated from the NN are presented and compared to the results obtained with the map-based technique discussed in Sec. \ref{sec:truthtagging}. Three main aspects are discussed: the modeling of single-jet distributions after jet weighting, the capability of the NN technique to provide an unbiased estimation of \eventeff, and the independence of the GNN performance on the choice of the sample used for training. 
 
\begin{figure}

    \includegraphics[]{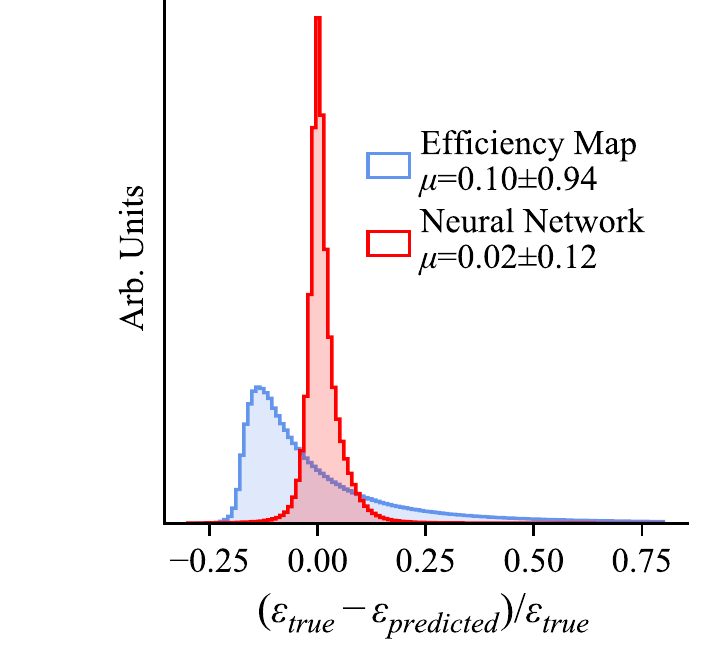}
   \label{fig:kinematics:dR}
    \includegraphics[]{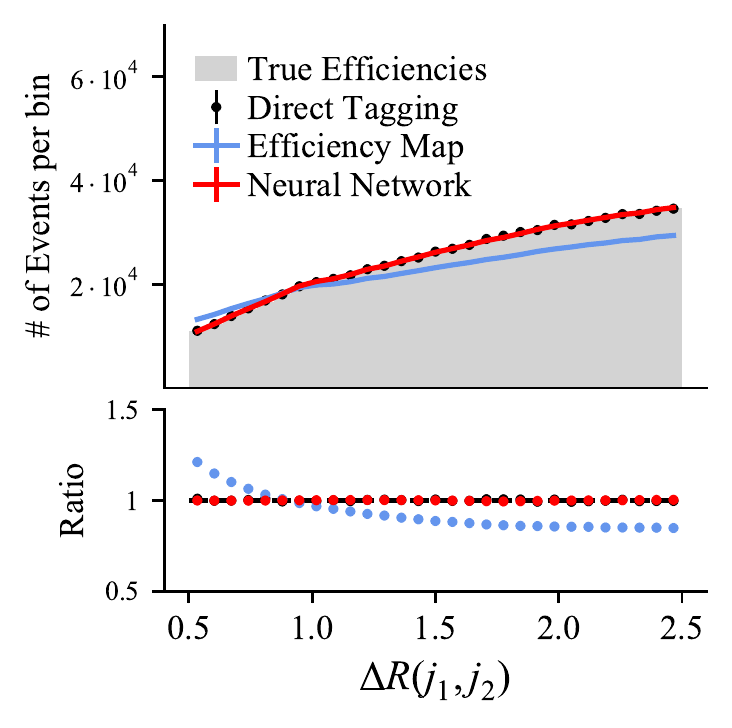}

    \label{fig:kinematics:pt}

\caption{Relative residuals distributions as predicted by the NN and the map-based approach for each individual jet in the event (top). The mean and RMS of the distributions are outlined in the plot. (bottom) Distribution of the jet  \deltar of the leading and subleading jet, obtained when the leading jet is classified as \btagged (black), compared to the same distributions obtained when jets are weighted with their true efficiency (grey), using the efficiency \tteff from the map-based approach (blue) or using the NN output (red). The lower pad shows the ratio between the two latter distributions and the one obtained with true weights.}
\label{fig:kinematics}
\end{figure}

Figure \ref{fig:kinematics} shows  the relative residuals $\frac{({\epsilon_{\textrm{true}}-\epsilon_{\textrm{predicted}}})}{{\epsilon_{\textrm{true}}}}$ for each jets in the event and \deltar distributions for the leading and subleading jet where the leading jet is classified as \btagged. $\epsilon_{\textrm{true}}$ is computed from Eq. \ref{eq:trueeff_jet}.  The results of \emph{direct tagging}\footnote{Only jets passing the \btagging classification are included in distributions, without any additional weight.} are also shown together with the jets weighted with either the predicted per-jet efficiency from the map-based (Eq~\ref{eq:truthtagging}) or NN approaches (Eq~\ref{eq:pratio}). While, as expected, the map-based approach is unable to provide good modeling of the \deltar distribution, the NN predictions are in good agreement with the distributions obtained with direct tagging and with true efficiency weights. These results give us confidence about the ability of the NN to approximate the set of relevant variables $\boldsymbol{\theta}$ as well as their dependency on the $\jeteff(\boldsymbol{\theta})$ of the different jets.

Results of the reweighing procedure are further studied when both the leading and sub-leading jets are classified as \bjets, and compared to those from direct tagging. In this case, the event weight is simply computed as the product of the efficiencies of \btagging each of the two jets, $\eventeff = \epsilon_{1}  \cdot\epsilon_{2}$. It is therefore important to study the modeling of distributions that capture correlations among individual jet observables, once event weights are applied.

\begin{figure*}
\centering
\includegraphics[]{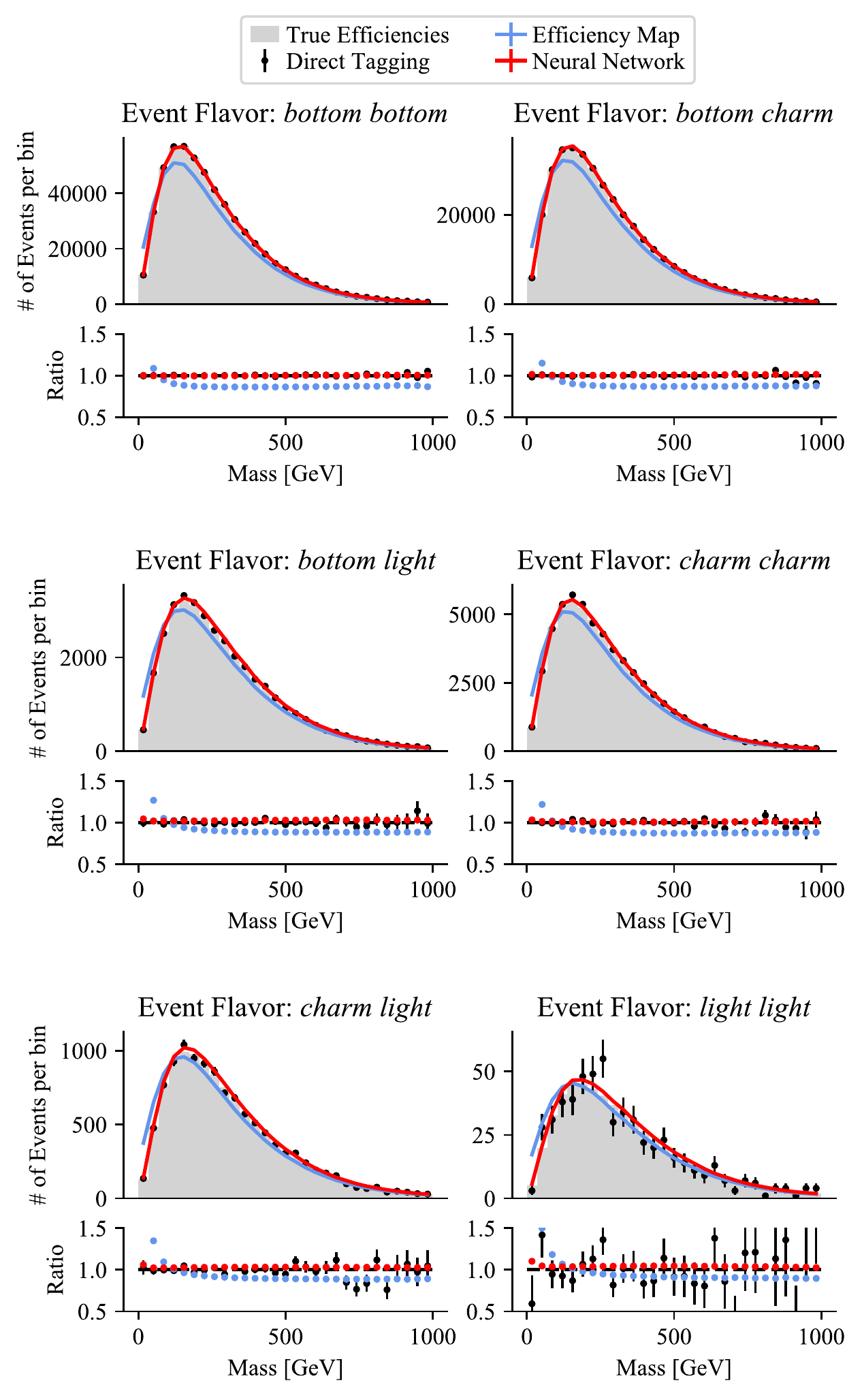}

\caption{Distribution of the invariant mass of the two leading jets, when the events are weighted by the product of true efficiencies, as calculated in Eq. \ref{eq:trueeff_jet} (grey). Also shown is the distribution for events where both jets are \btagged (direct tagging), or when the events are weighted using the estimated efficiency \tteff from the map-based approach (blue) or using the NN output (red). The lower pad shows the ratio between all distributions and the one obtained with true weights. Events are split into categories based on the true flavor of the two leading jets.
}\label{fig:mass_2}
\end{figure*}

The invariant mass distribution computed from the leading and subleading jets in each event is shown in Figure \ref{fig:mass_2}. The figures are further sub-divided based on the true flavors of the two jets. 
Similarly to the single-jet case, the NN predictions show good agreement compared to the true efficiency while the map-based approach is unable to properly capture the effect of close-by jets on \btagging. It can also be noted that the reweighing procedure based on NN predictions improves significantly the statistical uncertainly compared to the direct tagging.

\begin{figure}
    \includegraphics[]{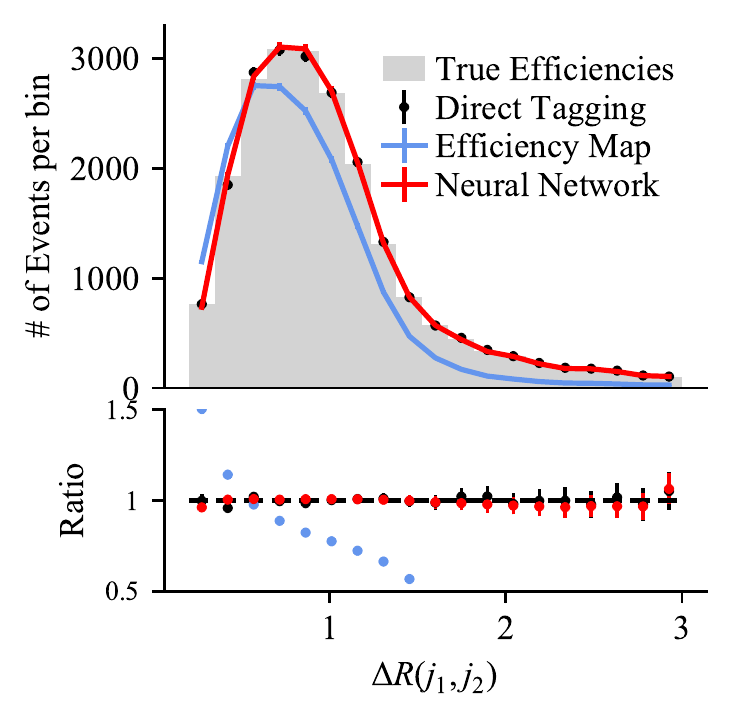}
    \includegraphics[]{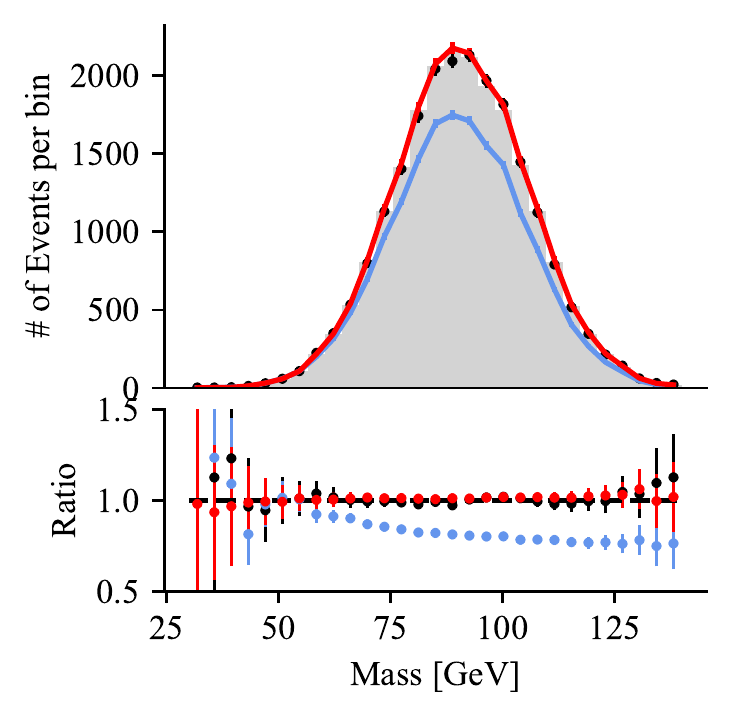}
\caption{Distribution of the \deltar (top) and invariant mass (bottom) of the leading-subleading jet system, obtained for events where these jets are classified as \btagged (blue), compared to the same distributions obtained when these jets are instead weighted with their probability of passing \btagging, calculated using the true weight $\epsilon$ from Eq. \ref{eq:trueeff_jet} (grey), using the efficiency \tteff from the map-based approach (blue) or using the NN output (red). The lower pad shows the ratio between the two latter distributions and the one obtained with true weights.}
\label{fig:Zsample}
\end{figure}

Finally, the generality of the method is probed by using the same network to reweight events from a separate sample with different jet \pt, $\eta$ and \deltar distributions compared to the training sample. For this purpose, events were simulated in which a boosted scalar particle decays in exactly two jets per event, where the \pt of the decaying particle is generated from an exponentially decaying distribution, and its mass is generated from a Gaussian distribution peaked at 90 GeV. The boson decays with a rate of 33\% to light-, $c$- or $b$-jets. Figure \ref{fig:Zsample} shows the results for the angular separation between the two decay products as well as for the reconstructed invariant mass of the generated boson. An overall good agreement is found between the NN results and direct tagging, similarly to the previous cases. This gives confidence about the universality of the proposed approach: as long as the phase space is sampled adequately during training, the efficiency estimated using the neural network is expected to be independent on the chosen sample.

\section{Discussion}
\label{sec:discussions}
In this section we summarize some of the main considerations aimed at generalizing the proposed approach for use cases beyond the toy model presented in this paper.
\begin{description}
    \item [\textbf{The size of $\boldsymbol{\theta}$:}] We used a relatively small number of variables that control the efficiency and required the network to only infer the variable $\deltar$. In real-life applications, $\boldsymbol{\theta}$ may include more variables and the related inference may be more complex in higher dimensions. To cope with this, the inputs variables $\mathbf{\Theta}$ needs to be enlarged using additional variables. Neural networks are a particularly suitable tool to perform this task due to their flexibility to cope with higher dimensions. Any variables potentially correlated with the tagging decision could be used to ensure that all correlations are captured.
    
    \item [\textbf{The functional form of $\epsilon(\boldsymbol{\theta})$:}] We assumed a relatively simple efficiency in Eq.~\ref{eq:trueeff_jet}. In principle, the neural network can learn any function, no matter how complex the functional form is, as shown in  Ref.~\cite{sets}. The method can be used  in scenarios where the form of $\epsilon(\boldsymbol{\theta})$ may present more complex dependencies between the efficiency and the relevant variables $\boldsymbol{\theta}$. 
    
    \item [\textbf{Systematic uncertainties:}] In the applications of the simple efficiency maps, the insufficient capture of the existing underlying correlations requires the introduction of systematic uncertainty. This method is aimed at avoiding this systematic error, it will, however, require thorough checks to ensure that its estimates are accurate.
    
    

    \item [\textbf{Generalization of the method:}] In the proposed approach we have focused our studies to approximate efficiency, i.e.  density ratios between two complementary classes. The method can also be generalized to approximate ratios between two separate classes\footnote{In such cases, the loss function needs to be changed to cope with non-complementary classes as discussed in Ref.~\cite{losstrick}}.  A multidimensional ratio between two classes could be used in a variety of different applications, such as to derive multi-dimensional scale factors from data to correct the tagging efficiency in Monte Carlo simulation.
\end{description}

\section{Conclusions}
\label{sec:conclusions}


The parametrization of classifier efficiencies can play an important role to mitigate the limitations in the number of simulated events at LHC experiments. To be effective, parametrized classifier efficiencies need to be accurate in any context and therefore need to capture the dependencies on event properties that are used in analyses and which entail variations of efficiencies.
A new technique that optimally exploits these dependencies is proposed. This technique is based on graph neural networks that provide an estimate of ratios between multidimensional local densities. We use the case of the identification of heavy-flavor jets as a topical example building a toy model based on ad-hoc parameterizations of the classifier efficiency inspired by the observed dependencies of \btagging performance in the ATLAS and CMS experiments. A Graph Neural Network is used to exploit correlations between jets in the event to provide an unbiased parametrization of the efficiency.

A toy example is used to probe the performance of the method, which
 takes as an input the true flavors and momenta of reconstructed jets, and returns the  \btagging efficiency of each. These efficiencies are used to build the per-event weights in a sample of simulated events with multiple $b$-tagged jets. We use the estimated efficiency for the event reweighing technique which is used to reduce the statistical fluctuations of Monte Carlo samples after classification. 

Results show good compatibility between per-jet and per-event kinematic distributions obtained with the proposed approach and the distributions expected from the direct application of \btagging. We also show that the proposed technique can generalize to samples with input distributions differing significantly compared to the training sample.

\section*{Acknowledgment}

EG and JS were supported by the NSF-BSF Grant 2017600
and the ISF Grant 2871/19. JS research was partially supported by the Israeli Council for Higher Education (CHE) via the Weizmann Data Science Research Center. 

CB, FADB, GF, MK and VI research was partially supported by the grant "Sviluppo di algoritmi innovativi di Deep Learning per dati altamente sparsificati e applicazione all'identificazione di particelle prodotte nei decadimenti del bosone di Higgs negli esperimenti a LHC".

\section*{Compliance with Ethical Standards}
On behalf of all authors, the corresponding authors
state that there is no confict of interest.


%
%


%
%



\begin{appendices}
\section{Sample generation details}
\label{app:samp}
This section describes the event generation of the toy model employed throughout this paper. At least two jets with $\pt>\SI{20}{GeV}$ and $|\eta|<2.0$ are generated. For each jet in the event, the jet transverse momentum is sampled from a gaussian distribution centered at 20 GeV with a width of 200 GeV, the sampling range is chosen to be [20, 600] GeV. The pseudo-rapidity of the leading jet in the event is sampled from a gaussian distribution centered at 0 with a width of 0.5 while the the azimuthal angle is sampled from a uniform distribution bounded in [0, 2$\pi$]. The angular variables of the other jets in the event are chosen by sampling from the square root of the angular distance, $\sqrt{\deltar}$, with $\deltar=\sqrt{(\eta_i-\eta_j)^2 + (\phi_i-\phi_j)^2}$  computed w.r.t. the leading jet. For a given value of $\deltar$, the jet angles are  sampled from a uniform distribution in the $\eta-\phi$ plane at the fixed \deltar value. The masses of the single jets are fixed at 2 GeV. These parameters ensure an invariant mass distribution similar to the one obtained in \WZjets events, as mentioned in the main body of the paper.

\section{Model Architecture}
\label{app:MA}
\begin{figure*}
\centering
\includegraphics[]{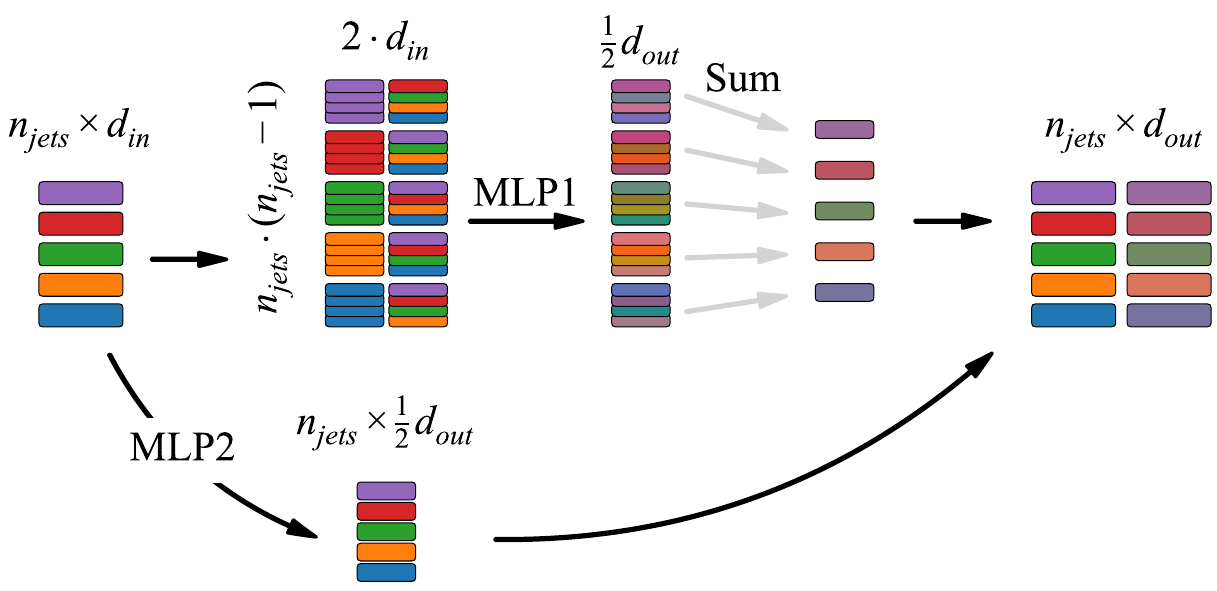}
\caption{GN block architecture.}
\label{fig:gnn_arch}
\end{figure*}

\paragraph{GNN Architecture.} 
The GNN is built from a stack of "GN blocks" as described in~\cite{GNN2}. The GN block is shown schematically Figure \ref{fig:gnn_arch}. 

 Each GN block takes in a matrix with shape  $n_{jets}\times d_{in}$, where $d_{in}$ is the size of the vector representing each jet. The output is a $n_{jets}\times d_{out}$ matrix where each jet representation has been updated based on the representation of the other jets in the event.
 
 Internally, the output representation is formed from a concatenation of two components. 
 
 The first component is a jet representation created by collecting information from other jets - first the input is rearranged to form all the ordered pairs of jets ($n\cdot(n-1)$ for $n$ jets in an event) by concatenating the input features of the two jets. A MLP  is then applied to the jet-pairs (MLP1 in Figure~\ref{fig:gnn_arch}). The output is summed for groups of jet-pairs who share the same "first jet" (note the pairs are ordered), resulting in a representation of size $\frac{1}{2} d_{out}$ for each of the $n_{jets}$. This representation is passed through another MLP (MLP3, not shown in Figure~\ref{fig:gnn_arch}), which maintains the same output size.
 
 The second component is formed by an MLP (MLP2 in Figure~\ref{fig:gnn_arch}) applied to each jet, creating a representation of size $\frac{1}{2} d_{out}$. 
 
 The resulting $n_{jets}\times d_{out}$ representation is normalized, such that each jet representation has Euclidean norm of 1.
 
 The GN blocks are applied to the input data sequentially. After the application of each GN block, the initial input of size $n_{jets}\times \text{jet features}$ is concatenated with the output (a "skip connection").
 
\subsection*{Model Details}

GNN layer sizes ($d_{in}$,$d_{out}$):
\begin{itemize}
    \item $(4,256)$
    \item 3 layers of $(256+4,256)$
\end{itemize}

GN block MLP1:
ReLU activation between each layer, and a final Tanh activation on the final layer.
\begin{itemize}
    \item $(2\cdot d_{in},\frac{1}{2}\cdot(2\cdot d_{in}+\frac{1}{2}\cdot d_{out}))$
    \item $(\frac{1}{2}\cdot(2\cdot d_{in}+\frac{1}{2}\cdot d_{out}),\frac{1}{2}\cdot(2\cdot d_{in}+\frac{1}{2}\cdot d_{out}))$
    \item $(\frac{1}{2}\cdot(2\cdot d_{in}+\frac{1}{2}\cdot d_{out}),\frac{1}{2}d_{out})$
\end{itemize}
GN block MLP2:
ReLU activation between each layer, and a final Tanh activation on the final layer.
\begin{itemize}
    \item $(d_{in},\frac{1}{2}\cdot( d_{in}+\frac{1}{2}\cdot d_{out}))$
    \item $(\frac{1}{2}\cdot( d_{in}+\frac{1}{2}\cdot d_{out}),\frac{1}{2}d_{out})$
\end{itemize}
GN block MLP3:
ReLU activation between each layer, and a final Tanh activation on the final layer.
\begin{itemize}
    \item $(\frac{1}{2}d_{out},\frac{1}{2}d_{out})$
    \item $(\frac{1}{2}d_{out},\frac{1}{2}d_{out})$
\end{itemize}
Jet Efficiency MLP layers ($d_{in}$,$d_{out}$):
\begin{itemize}
    \item $(256+4,256)$
    \item $(256,128)$
    \item $(128,50)$
    \item $(50,1)$
\end{itemize}

\end{appendices}

\end{document}